\documentclass[sigconf,screen,nonacm,review=false,timestamp=false]{acmart}

\AtBeginDocument{%
  \providecommand\BibTeX{{%
    \normalfont B\kern-0.5em{\scshape i\kern-0.25em b}\kern-0.8em\TeX}}}

\setcopyright{acmcopyright}
\copyrightyear{2018}
\acmYear{2018}
\acmDOI{XXXXXXX.XXXXXXX}

\acmConference[Conference acronym 'XX]{Make sure to enter the correct
  conference title from your rights confirmation emai}{June 03--05,
  2018}{Woodstock, NY}
\acmPrice{15.00}
\acmISBN{978-1-4503-XXXX-X/18/06}

\begin{document}

\title{Soft-Search: Two Datasets to Study the Identification and Production of Research Software}

\author{Eva Maxfield Brown}
\email{evamxb@uw.edu}
\orcid{0000-0003-2564-0373}
\affiliation{%
  \institution{University of Washington Information School}
  \city{Seattle}
  \state{Washington}
  \country{USA}
}

\author{Lindsey Schwartz}
\email{lsschwar@uw.edu}
\orcid{0000-0001-8338-0288}
\affiliation{%
  \institution{University of Washington Information School}
  \city{Seattle}
  \state{Washington}
  \country{USA}
}

\author{Richard Lewei Huang}
\email{lwhuang@uw.edu}
\orcid{0000-0002-0264-9300}
\affiliation{%
  \institution{University of Washington Information School}
  \city{Seattle}
  \state{Washington}
  \country{USA}
}

\author{Nicholas Weber}
\email{mweber@uw.edu}
\orcid{0000-0002-6008-3763}
\affiliation{%
  \institution{University of Washington Information School}
  \city{Seattle}
  \state{Washington}
  \country{USA}
}

\renewcommand{\shortauthors}{Brown, et al.}

\begin{abstract}
Software is an important tool for scholarly work, but software produced
for research is in many cases not easily identifiable or discoverable. A
potential first step in linking research and software is software
identification. In this paper we present two datasets to study the
identification and production of research software. The first dataset
contains almost 1000 human labeled annotations of software production
from National Science Foundation (NSF) awarded research projects. We use
this dataset to train models that predict software production. Our
second dataset is created by applying the trained predictive models
across the abstracts and project outcomes reports for all NSF funded
projects between the years of 2010 and 2023. The result is an inferred
dataset of software production for over 150,000 NSF awards. We release
the Soft-Search dataset to aid in identifying and understanding research
software production: https://github.com/si2-urssi/eager    
\end{abstract}



\maketitle

\setlength{\parskip}{-0.1pt}

\hypertarget{introduction}{%
\section{Introduction}\label{introduction}}

Software production, use, and reuse is an increasingly crucial part of
scholarly work
\citep{open_source_code_repo_predict_impact, Trisovic2021ALS}. While
historically underutilized, citing and referencing software used during
the course of research is becoming common with new standards for
software citation \citep{Katz2021RecognizingTV, Du2022UnderstandingPI}
and work in extracting software references in existing literature
\citep{cz_software_mentions}. However, records of software production
are not readily identifiable or available at scale in the way that
peer-reviewed publications or other scholarly outputs are
\citep{Schindler2022TheRO}. To make progress on this problem, we
introduce two related datasets for studying and inferring software
produced as a part of research, which we refer to as the Soft-Search
dataset.

The Soft-Search dataset is aimed at identifying research projects which
are likely to have produced software while funded by a federal grant. We
start by identifying GitHub repositories that acknowledge funding from
at least one National Science Foundation (NSF) award. We then annotate
each GitHub repository found with a binary decision for its contents:
software or not-software (e.g.~not all github repositories contain
software, they might include research notes, course materials, etc.). We
then link each annotated GitHub repository to the specific NSF award
ID(s) referenced in its README.md file. Finally, we compile the
Soft-Search Training dataset using the annotations for each GitHub
repository, and the text from the linked NSF award abstract and the
project outcomes report.

Using the Soft-Search Training dataset, we train a variety of models to
predict software production using either the NSF award abstract or
project outcomes report text as input. We use the best performing models
to then infer software production against all awards funded by the
National Science Foundation from 2010 to 2023 (additional details are
offered in Section~\ref{sec-data-collection}). The predictions and
metadata for each NSF award between the 2010 and 2023 period are
compiled to form the Soft-Search Inferred dataset.

In total, our new Soft-Search dataset includes the following
contributions:

\begin{enumerate}
\def\labelenumi{\arabic{enumi}.}
\item
  Soft-Search Training: A ground truth dataset compiled using linked NSF
  awards and GitHub repositories which have been annotated for software
  production.
\item
  Multiple classifiers which infer software production from either the
  text of an NSF award's abstract or project outcomes report.
\item
  Soft-Search Inferred: A dataset of more than 150,000 NSF funded awards
  from between 2010 and 2023. Each award has two predictions for
  software production: one from prediction using the abstract text and
  the other from prediction using the project outcomes report text.
\end{enumerate}

The rest of the paper proceeds as follows. In
Section~\ref{sec-data-collection} we detail the data collection and
annotation process used for creating the Soft-Search Training dataset.
In Section~\ref{sec-models} we briefly describe the model training
process and report results. In Section~\ref{sec-soft-search-dataset} we
provide summary statistics for the Soft-Search Inferred dataset and
observe trends in software production over time. We conclude with
discussion regarding the limitations of our approach and opportunities
for future work.

\hypertarget{sec-data-collection}{%
\section{Data Collection and Annotation}\label{sec-data-collection}}

\hypertarget{sec-finding-soft}{%
\subsection{Finding Software Produced by NSF
Awards}\label{sec-finding-soft}}

The first step in our data collection process was to find software
outputs from National Science Foundation (NSF) funded research. This
step has two potential approaches. The first approach is a manual search
for references and promises of software production within NSF award
abstracts, project outcome reports, and papers supported by each award.
This first approach is labor intensive and may be prone to labeling
errors because while there may be a promise of software production in
these documents, it may not be possible to verify such software was
ultimately produced. The other approach is to predict software
production using a trained model. We pursue this approach with the
caveat that there are also potential label errors.

To gather examples of verifiable software production, we created a
Python script which used the GitHub API to search for repositories which
included reference to financial support from an NSF award in the
repositories README.md file. Specifically our script queried for
README.md files which contained any of the following text snippets:
`National Science Foundation', `NSF Award', `NSF Grant', `Supported by
NSF', or `Supported by the NSF'. GitHub was selected as the basis for
our search because of its widespread adoption and mention in scholarly
publication \citep{riseofgithubinscholarlypublication}. This search
found 1520 unique repositories which contained a reference to the NSF in
the repository's README.md file.

\hypertarget{software-production-annotation}{%
\subsection{Software Production
Annotation}\label{software-production-annotation}}

The next step in our data collection process was to annotate each of the
GitHub repositories found as either ``software'' or ``not software.'' In
our initial review of the repositories we had collected, we found that
the content of repositories ranged from documentation, experimental
notes, course materials, collections of one-off scripts written during a
research project, to more typical software libraries with installation
instructions, testing, and community support and use.

Using existing definitions of what constitutes research software to form
the basis of our annotation criteria
\citep{martinez_ortiz_carlos_2022_7185371, sochat2022research}, we
conducted multiple rounds of trial coding on samples of the data.
Fleiss' kappa was used to determine if there was agreement between our
research team on whether ten GitHub repositories contained `software' or
not. On each round of trial coding ten GitHub repositories were randomly
selected from our dataset for each member of our research team to
annotate independently. When assessing a repository, members of the
research team were allowed to use any information in the repository to
determine their annotation (i.e.~the content of the README.md file, the
repository activity, documentation availability, etc.)

Our final round of trial coding showed that there was near perfect
agreement between the research team (K=0.892)
\citep{viera2005understanding}.

Our final annotation criteria was generally inclusive of labeling
repositories as software, rather there were specific exclusion criteria
that resulted in a repository being labeled as ``not software''.
Specifically repositories were labeled as ``not software'' when a
repository primarily consisted of:

\begin{enumerate}
\def\labelenumi{\arabic{enumi}.}
\item
  project documentation or research notes
\item
  teaching materials for a workshop or course
\item
  the source code for a project or research lab website
\item
  collections of scripts specific to the analysis of a single experiment
  without regard to further generalizability
\item
  utility functions for accessing data without providing any additional
  processing capacity
\end{enumerate}

We then annotated all GitHub repositories in our dataset as either
``software'' or ``not software'' according to our agreed upon annotation
criteria.

\hypertarget{linking-github-repositories-to-nsf-awards}{%
\subsection{Linking GitHub Repositories to NSF
Awards}\label{linking-github-repositories-to-nsf-awards}}

Our final step in the data collection process was to link the annotated
GitHub repositories back to specific NSF awards. To do so, we created a
script which would load the webpage for each GitHub repository, scrape
the content of the repository's README and find the specific NSF award
ID number(s) referenced. While annotating the dataset, and with this
script, our dataset size was reduced as we found some repositories were
returned in the initial search because of the ``NSF'' acronym being used
by other, non-United-States governmental agencies which also fund
research.

When processing each repository, our Python script would load the README
content, search for NSF Award ID patterns with regular expressions, and
then verify that each NSF award ID found was valid by requesting
metadata for the award from the NSF award API.

We then retrieved the text for each award's abstract and project
outcomes report. This was the final step of our data collection process
and allowed us to create a dataset of 446 unique NSF awards labeled as
`produced software' and 471 unique NSF awards labeled as `did not
produce software'.

\hypertarget{sec-models}{%
\section{Predictive Models}\label{sec-models}}

Using the compiled Soft-Search Training dataset, we trained three
different models using the text from either the award abstract or
project outcomes report. The models trained include a logistic
regression model trained with TF-IDF word embeddings
(\texttt{tfidf-logit}), a logistic regression model trained with
semantic embeddings (\texttt{semantic-logit}), and a fine-tuned
transformer (\texttt{transformer}). The semantic embeddings and the base
model from which we fine-tuned our own transformer model was the
`distilbert-base-uncased-finetuned-sst-2-english' model
\citep{hf_canonical_model_maintainers_2022}. Each model was trained with
80\% of the Soft-Search Training dataset. We then test each of the
models and use F1 to rank each model's performance.

\hypertarget{tbl-model-results-from-abstract}{}
\begin{table}
\caption{\label{tbl-model-results-from-abstract}Predictive Model Results (Trained with Abstract Text) }\tabularnewline

\centering
\begin{tabular}{llrrrr}
\toprule
{} &           model &  accuracy &  precision &  recall &     f1 \\
\midrule
0 &     tfidf-logit &     0.674 &      0.674 &   0.674 &  0.673 \\
1 &     transformer &     0.636 &      0.608 &   0.697 &  0.649 \\
2 &  semantic-logit &     0.630 &      0.630 &   0.630 &  0.630 \\
3 &           regex &     0.516 &      0.515 &   0.516 &  0.514 \\
\bottomrule
\end{tabular}
\end{table}

Table~\ref{tbl-model-results-from-abstract} reports the results from
training using the abstract text as input. The best performing model was
the \texttt{tfidf-logit} which achieved an F1 of 0.673.

\hypertarget{tbl-model-results-from-project-outcomes}{}
\begin{table}
\caption{\label{tbl-model-results-from-project-outcomes}Predictive Model Results (Trained with Project Outcomes Report Text) }\tabularnewline

\centering
\begin{tabular}{llrrrr}
\toprule
{} &           model &  accuracy &  precision &  recall &     f1 \\
\midrule
0 &     tfidf-logit &     0.745 &      0.745 &   0.745 &  0.745 \\
1 &     transformer &     0.673 &      0.638 &   0.771 &  0.698 \\
2 &  semantic-logit &     0.633 &      0.633 &   0.633 &  0.632 \\
3 &           regex &     0.510 &      0.507 &   0.510 &  0.482 \\
\bottomrule
\end{tabular}
\end{table}

Table~\ref{tbl-model-results-from-project-outcomes} reports the results
from training using the project outcomes reports as input. The best
performing model was the \texttt{tfidf-logit} which achieved an F1 of
0.745.

While the models trained with the project outcomes reports were trained
with less data, the best model of the group achieved a higher F1 than
any of the models trained with the abstracts. While we have not
investigated further, we believe this to be because the project outcomes
reports contain more direct citation of produced software rather than an
abstract's promise of software production.

The data used for training, and functions to reproduce these models, are
made available via our Python package:
\href{https://github.com/si2-urssi/eager}{\texttt{soft-search}}.

\hypertarget{sec-soft-search-dataset}{%
\section{The Soft-Search Dataset}\label{sec-soft-search-dataset}}

Using the predictive models, we compile the Soft-Search Inferred dataset
which contains the metadata, abstract text, and project outcomes report
text, for all NSF awarded projects during the 2010-2023 period. The
Soft-Search Inferred dataset additionally contains our predictions for
software production using both texts respectively.

\hypertarget{tbl-soft-search-stats}{}
\begin{table}
\caption{\label{tbl-soft-search-stats}Composition of the NSF Soft Search Dataset }\tabularnewline

\centering
\begin{tabular}{llrrr}
\toprule
{} & Program &  \# Awards &  \# Software &  \% Software \\
\midrule
0 &     MPS &     32885 &       19178 &    0.583184 \\
1 &    CISE &     24633 &       13274 &    0.538871 \\
2 &     ENG &     22900 &       11242 &    0.490917 \\
3 &     GEO &     17822 &        5142 &    0.288520 \\
4 &     BIO &     16990 &        6013 &    0.353914 \\
5 &     EHR &     13703 &         575 &    0.041962 \\
6 &     SBE &     13318 &        1966 &    0.147620 \\
7 &     TIP &      8597 &        4501 &    0.523555 \\
8 &    OISE &      2329 &         636 &    0.273079 \\
9 &     OIA &       498 &         123 &    0.246988 \\
\bottomrule
\end{tabular}
\end{table}

\hypertarget{trends-and-observations}{%
\subsection{Trends and Observations}\label{trends-and-observations}}

\begin{figure}

{\centering \includegraphics[width=\linewidth]{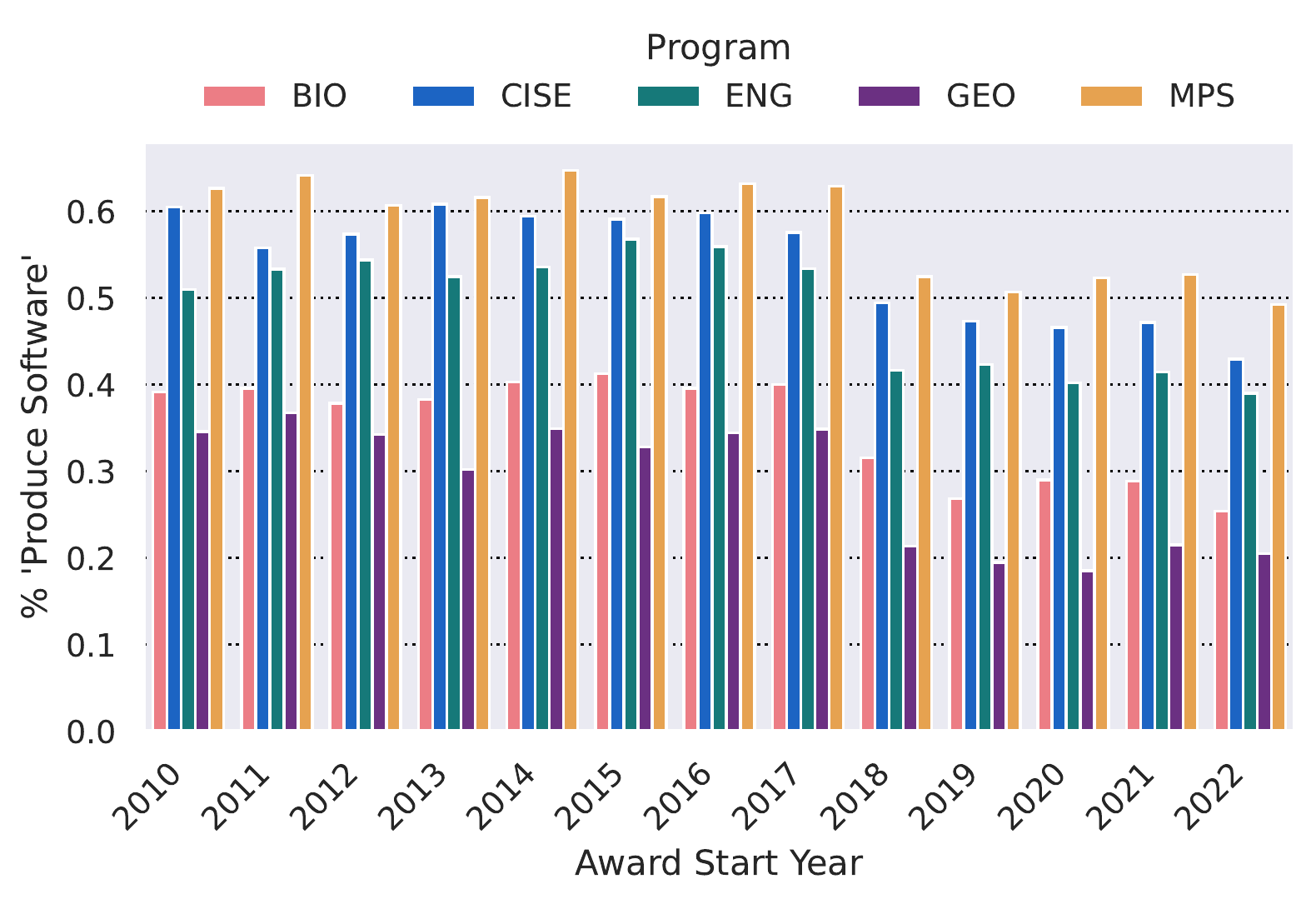}

}

\caption{\label{fig-soft-over-time}Software Production Over Time (Using
Predictions from Abstracts)}

\end{figure}

Using the Soft-Search Inferred dataset we can observe trends in software
production over time. Figure~\ref{fig-soft-over-time} plots the percent
of awards which we predict to have produced software (using the award's
abstract) over time. While there are minor year-to-year deviations in
predicted software production, we observe the ``Math and Physical
Sciences'' (MPS) funding program as funding the most awards which we
predict to produce software, with ``Computer and Information Science and
Engineering'' (CISE), and ``Engineering'' (ENG) close behind.

\begin{figure}

{\centering \includegraphics[width=\linewidth]{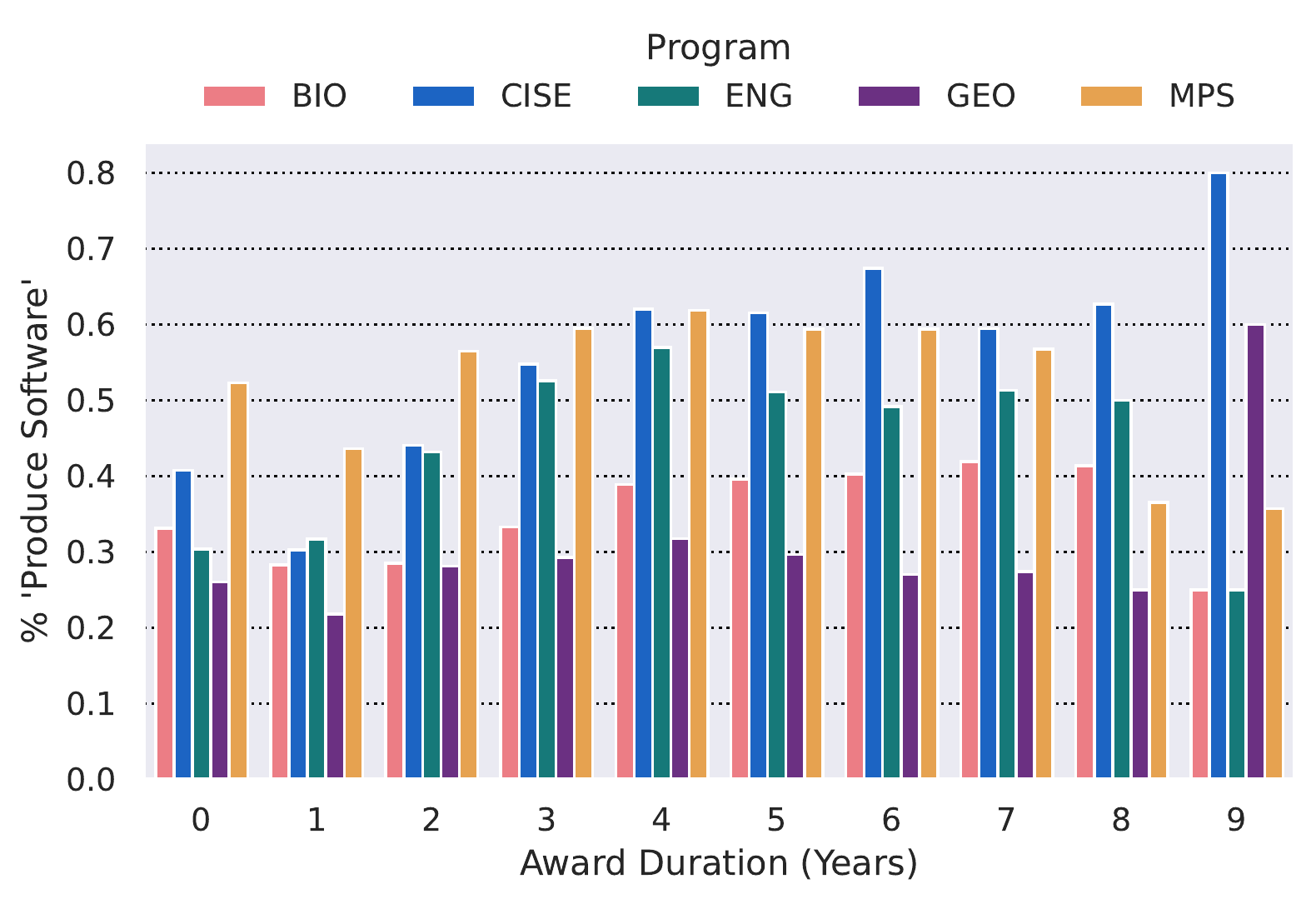}

}

\caption{\label{fig-soft-over-duration}Software Production Grouped By
Award Duration (Using Predictions from Abstracts)}

\end{figure}

We can additionally observe trends in software production as award
duration increases. Figure~\ref{fig-soft-over-duration} plots the
percent of awards which we predict to have produced software (using the
award's abstract) grouped by the award duration in years. We note that
as award duration increases, the percentage of awards which are
predicted to have produced software also tends to increase.

\hypertarget{conclusion}{%
\section{Conclusion}\label{conclusion}}

We introduce Soft-Search, a pair of novel datasets for studying software
production from NSF funded projects. The Soft-Search Training dataset is
a human-labeled dataset with almost 1000 examples used to train models
which predict software production from either the NSF award abstract
text or the project outcomes report text. We used these models to
generate the Soft-Search Inferred dataset. The Soft-Search Inferred
dataset includes project metadata, the awards abstract and project
outcomes report, and predictions of software production for each NSF
funded project between 2010 and 2023. We hope that Soft-Search helps
further new studies and findings in understanding the role software
development plays in scholarly publication.

All datasets and predictive models produced by this work are available
from our GitHub repository:
\href{https://github.com/si2-urssi/eager}{\texttt{si2-urssi/eager}}.

\hypertarget{limitations}{%
\subsection{Limitations}\label{limitations}}

As discussed in Section~\ref{sec-data-collection}, the Soft-Search
Training dataset was entirely composed of NSF awards which ultimately
released or hosted software (and other research products) on GitHub. Due
to our data collection strategy, it is possible that each of the
predictive models learned not to predict if an NSF award would produce
software, but rather, if an NSF award would produce software hosted on
GitHub.

\hypertarget{future-work}{%
\subsection{Future Work}\label{future-work}}

As discussed in Section~\ref{sec-finding-soft}, our initial method for
attempting to find research software produced from NSF supported awards
was to search for references and promises of software production in the
abstract, project outcomes report, and attached papers of each award.
While attempting this approach to create the dataset, we found that many
awards and papers that reference computational methods do not provide a
reference web link to their code repositories or websites. In some
cases, we found repositories related to an award or paper via Google and
GitHub search ourselves. While we support including references to code
repositories in award abstracts, outcomes reports, and papers, future
research should be conducted on how to enable automatic reconnection of
papers and their software outputs.

\hypertarget{acknowledgements}{%
\section{Acknowledgements}\label{acknowledgements}}

We thank the USRSSI team, especially Karthik Ram for their input. This
material is based upon work supported by the National Science Foundation
under Grant 2211275.

\bibliographystyle{ACM-Reference-Format}
\bibliography{sample-base.bib}


\end{document}